\documentstyle[12pt]{article}
\begin{document}
\title{BUBBLE NUCLEATION IN \(\varphi^{4}\) MODELS AT ALL TEMPERATURES}
\author{Antonio Ferrera\\
        Physics Department, Tufts University, Medford, MA 02155\\}
\date{}
\maketitle
\begin{abstract}
One possible way in which phase transitions in the early universe may have
ocurred is via nucleation of bubbles of the new phase (true vacuum) in
the old phase (false vacuum). The technique
most widely used to compute the probability of bubble nucleation is based on
instanton methods in the context of the semiclassical approximation. At zero
temperature in \(3+1\) dimensions the nucleation rate is dominated by the O(4)
symmetric instanton, a sphere of radius $R$, while at temperatures
$T >> R^{-1}$, the decay is dominated by a ``cylindrical'' (static)
instanton wtih O(3) invariance. There has been discussion in the literature as
to whether the transition between these two regimens would be first order
(discontinuity in the first derivative of the nucleation rate at the transition
 temperature $T_{c}$), or second order (continuity of the first derivative, but
 discontinuity of the second derivative at $T_{c}$). In this paper we obtain
the finite temperature solutions corresponding to the quantum and the thermal
regimes, and compute their action as a
function of the temperature for different values of the wall thickness in a
$\varphi^{4}$ potential. Our
results indicate that only for the cases of very large wall thickness a second
order transition takes place, while for all the other cases a first order
transition occurs.\\
\end{abstract}
\baselineskip=24pt
\vspace{2em}

\section{Introduction}
The problem of the decay of a metastable state via quantum tunneling has
important applications in many branches of physics, from condesed matter to
particle physics or cosmology. In many cases however, the difficulties
associated with a full quantum treatment, and the fact that tunneling is a
non-perturbative effect, make it impossible for us to be able to make
quantitaive predictions about the behaviour of the system. When the nature of
the tunneling system allows the use of the semiclassical approximation though,
instanton techniques, used to describe quantum tunneling in that regime,
provide us with a formalism capable of producing accurate values for decay
rates. In condesed matter this formalism has
been applied to an array of problems ranging from the tunneling of an Abrikosov
vortex line out of a pinning potential in a superconductor [1-3], to the change
in the magnetization direction of thin films under field inversion [4]. In the
context of cosmology, perhaps the most compelling example is that of the study
of first order phase transitions in the early universe [5], although other
problems such as the decay of metastable topological defects formed during
these transitions [6], have also shown to be amenable to this sort of
treatment.\\
\\
   The basic tenet of the formalism is that the decay rate per unit time
(and where applicable per unit volume) is given by an expression of the form
\begin{equation}
\Gamma=Ae^{-S_{E}}
\end{equation}
where $S_{E}$ is the Euclidean action of the instanton: the classical solution
to the equations of motion under the potential barrier in euclidean time with
periodic boundary conditions. The instanton has turning points at the
configurations at which the system enters and
exits the barrier, and its analytic continuation to Lorentzian time at the
exit point describes the appearance of the system at the other side of the
barrier and its subsequent evolution. When the solution for the given boundary
conditions is not unique, one should take into account only the one with the
lowest $S_{E}$, which will dominate (1) (unless of course all or some of the
solutions happen to have
comparable actions, in which case one should take all relevant contributions
to $\Gamma$ into account). The determinantal prefactor $A$ in Eq.(1),
meanwhile, results from a gaussian functional integration over small
fluctuations around the
instanton. The formalism was first developed and applied to false vacuum decay
at zero temperature in refs [7-9]. The dominant solution in this case is a
maximally symmetric instanton
invariant under O(4) (for three spatial dimensions) [9].\\
\\
   In the context of quantum mechanics (i.e. \(0+1\)
dimensions) the formalism was extended to include finite temperature effects
by Affleck [10]. Under certain assumptions for the shape of the barrier he
found
that above a critical temperature determined by the curvature of the potential
at the top of the barrier, the dominant solution was a ``static'' (euclidean
time independent) instanton sitting at the bottom of the inverted potential
well, the decay rate produced by this solution recovering the expression for
purely classical thermal activation over the barrier that we would expect to
dominate the procces at high temperatures. The transition between the thermal
and quantum
regimens is dominated by solutions with a finite period in the Euclidean time
that smoothly interpolate between the zero temperature instanton and the
``static'' high temperature solution. The transition is thus a second order one
(i.e., the decay rate $\Gamma$ and its first derivative with respect to the
temperature are continous at the critical temperature of the transition,
$T_{c}$, but not so the second derivative).

   Chudnovsky [11] however showed that this situation is not generic, and that
the type of transition will in general be determined by $d\tau_{p}/dE_{eucl}$,
the derivative of the euclidean period of the motion $\tau_{p}$ with respect to
its euclidean energy. For the case in which the period vs. energy curve
possesses a minimum at values of $(E_{eucl},\tau_{p})$ different from those
associated with small oscillations around the minimum of the potential, it can
be shown that, at the transition temperature, the dominant solution changes
discontinously to the high temperature instanton. Consequently the curves for
the actions of both regimens cross each other at that point, and the derivative
of $S_{min}$ with respect to the temperature (and thus that of $\Gamma$)
changes
discontinuosly as we go from the thermal activation regime to
the quantum regime. This is what Chudnovsky called a {\em first order}
transition.  In this case we should be able to observe quite a sharp
change in the behaviour of $\Gamma$ as we go through the critical
region.\\
\\
   In the context of field theory very little work has been done
towards examining the structure of the intermediate regime instantons as well
as that of the transition between the two extreme cases of high and zero
temperature. Studying the bubble nucleation rate for phase transitions in the
early universe, Linde [5] suggested that periodic instantons would smoothly
interpolate between these two regimens, much in the same way that Affleck had
shown could be the case for quantum mechanics shortly before. In this picture
then, one would
have Coleman's O(4) invariant spherical bounce of radius $R$ dominating at
zero temperature. As one goes into finite temperature one would first
encounter a periodic array of, rather weakly interacting, widely separated
spheres, but after we have increased the temperature up to values
$T \sim (2R)^{-1}$ and beyond we no longer can fit the spheres into the
corresponding periods, and they would start merging into one another producing
 what has
become known as a ``wiggly cylinder'' solution. As we keep increasing the
temperature further, the wiggles eventually smoothly straighten out, and the
solution goes into the O(3) invariant ``cylinder'' that dominates the thermal
activation regime.

   In correspondence with the work of Chudnovsky in quantum mechanics however,
Garriga [12], in a recently published paper, has shown that within the thin
wall
approximation the transition is in fact not second order, but rather, first
order. The wiggly cylinder
solutions do indeed exist within some range of the temperature, but when they
exist they always have a higher action than any of the other solutions for the
same temperature, and therefore they never dominate the exponential in (1). The
instanton that saturates (1) goes then abruptly from being a periodic series of
separated bounces to being the cylindrical solution, and the rate of change of
the action with the temperature changes discontinuosly. The
transition thus produced is in consequence first order. It may be worth noting
at this point that the terminology used to
describe this transitions (i.e., first and second order) refers {\em only} to
the type of transition between the two different instanton regimes, and not to
the phase transition of the physical system itself. \\
\\
   In this paper we aim to explore the nature of the transition between the
thermal hopping and the zero temperature regime when we go beyond the thin wall
approximation, in the context of the bubble nucleation problem in $\varphi^{4}$
models just mentioned. In order to do this, we will
develop an algorithm to numerically compute the relevant finite temperature
solutions for a given value of $f$ (the parameter that governs the width of the
wall as compared to the radius of the bubble), and then compute the
corresponding curve for the action as a function of temperature. Finally we
repeat this procedure for different values of $f$, covering the most
significant physical cases. Our results indicate that in
the case of extremely thick wall (when the tunneling occurs over a very shallow
barrier) we do get a second order transition, but that in all other cases the
transition is in fact first order, in particular,close to the
thin wall regime our results agree rather well with those of Garriga. In spite
of the fact that in any realistic model of the phase transition the potential
will be temperature dependent, whereas we use temperature independent
potentials, we do not expect
 this to significantly change our results, as will be seen at the conclusion.
The paper is organized as follows: in section II we give a quantitative account
of the finite temperature transition in the spirit of [11], in section III we
present the results that Garriga
obtained for bubble nucleation in the thin wall limit, in section IV we
introduce our numeric algorithm and in section V we present our results.
Finally, conclusions are in section VI
\section{Finite Temperature Instantons in Quantum Mechanics}
   Acording to [7] the decay rate in the semiclassical limit ($S >> {\hbar}$)
is,
\begin{equation}
\Gamma{\propto}exp(-S_{min}/{\hbar})
\end{equation}
where $S_{min}(T)$ is evaluated along the $q(\tau)$ trajectory with period
$\tau_{p}=\beta={\hbar}/k_{B}T$ that minimizes the Euclidean time action
\begin{equation}
S(T)={\int}d\tau{\cal L}={\int}d{\tau}[\frac{1}{2}M{\dot{q}}^{2}+V(q)]\;\;\;,
\end{equation}
with $\dot{q}=dq/d\tau$ and where the integral is performed over a whole period
$\beta$ (see [7-9], [11] for details; from now on we will use $\beta$ to
designate both the
euclidean period and the usual inverse temperature). Such trajectories will of
course be solutions of the classical equations of motion in Euclidean time
\begin{equation}
M\ddot{q}=\frac{dV}{dq}
\end{equation}
consistent with the boundary conditions, that is, periodicity with period
$\beta$.  (i.e., $q_{1}=q(-\beta/2)=q(\beta/2)$ and $q_{2}=q(0)$ will be the
turning points of
the motion). By noting that (4) is the equation of motion of a particle in a
potential -V in Lorentzian time, we can draw two consequences: first, that the
(positive) energy
\begin{equation}
E(\beta)=V(q)-\frac{1}{2}M{\dot{q}}^{2}
\end{equation}
is conserved. And, second, that by inverting the barrier
we can recognize the types of solutions that we will have for the motion of the
particle in the well (see Fig. 1). It is
then immediately obvious that there are two solutions with period
$\beta$, one that actually is a time independent solution sitting at the
bottom of the well, $q=q_{0}$, and
another one that corresponds to periodic motion in the inverted well between
$q_{1}(E)$ and $q_{2}(E)$. For the first one we will have from (3) an action
\begin{equation}
S_{0}=V(q_{0}){\beta}\;\;\;,
\end{equation}
whereas for the second one, use of (5) into (3) yields
\begin{equation}
S_{T}=2(2M)^{1/2}{\int}_{q_{1}}^{q_{2}} dq[V(q)-E]^{1/2} + E\beta\;\;\;,
\end{equation}
where $E$ and $\beta$ are related through
\begin{equation}
{\beta}=(2M)^{1/2}{\int}_{q_{1}}^{q_{2}} dq[V(q)-E]^{-1/2}\;\;\;.
\end{equation}
Then, from (6), (7) and (8) one obtains
\begin{equation}
\frac{dS_{0}}{d{\beta}}=E_{0}\;,\;\;\;\;\;\;\;\frac{dS_{T}}{d{\beta}}=E >0
\;\;\;,
\end{equation}
and
\begin{equation}
\frac{d^{2}S_{0}}{d{\beta}^{2}}=0\;,\;\;\;\;\;\;\;\;
\frac{d^{2}S_{T}}{d{\beta}^{2}}=\frac{1}{d{\beta}/dE}\;\;\;.
\end{equation}
The behaviour of these derivatives at the temperature $T_{c}$ at which
$S_{T}=S_{0}$
will then determine the type of transition between the two regimens: if at
$T_{c}$ we have $S'_{0}{\neq}S'_{T}$ the transition will be first order, but if
the equality holds, we will have a second order transition for which
$S_{0}''{\neq}S_{T}''$ as we are about to see. From (10) however we see that
this will in turn be determined by the behaviour of $d\beta/dE$ (which
can be interpreted as the specific heat of the system).

   As a first example we can look at potentials of the form $-q^{2}+q^{3}$ or
$-q^{2}+q^{4}$ [10]. In this case $\beta$ monotonically decreases with $E$
(Fig.
2), until it eventually reaches $\beta_{0}$, the period of small oscillations
at the bottom of the potential. As seen from Fig.2,  $S_{T}{\rightarrow}S_{0}$
and $S_{T}'{\rightarrow}E_{0}=S_{0}'$  as  $\beta{\rightarrow}\beta_{0}$, or
what is the same, as $T{\rightarrow}T_{0}$, $T_{0} (={\hbar}/k_{B}\beta_{0})$
being the transition temperature in this case. The second derivative $S_{T}''$
will however change discontinually as we go across $T_{0}$, and will, as a
matter of fact
be undefined at that temperature (of course a more careful quantum analysis is
needed at this point). This is what Chudnovsky called a second order
transition.

   The other case of interest to us is that for which $\beta(E)$ develops a
minimum, at say $(E_{1},\beta_{1})$, before reaching $\beta_{0}$ at $E_{0}$
(see
Fig.3). This will be the
case for potentials that change very slowly both at the top and the bottom
(although still being parabolic) but that are very steep in the middle. In this
case we see that, as $E$ decreases from $E_{0}$ to $E_{1}$, we
have $S_{T}'=E<E_{0}=S_{0}'$ and $S_{T}''=1/\beta'>0$. The corresponding
curve for $S_{T}(\beta)$ will then be a concave curve that stays always
above the
thermal activation curve. At $E_{1}$ however $\beta'=0$ and $S_{T}''$ is
undefined, and for $E<E_{1}$ we have to have a convex curve that quickly goes
into a straight horizontal line for high vlaues of $\beta$ (Fig. 3b).
$S_{T}$ then has
to cross over the $S_{0}$ curve at some period $\beta_{c}$. The derivatives of
the two curves will be different at that point and, therefore, the first
derivative of $S_{min}$ will suffer a discontinous change as we go across the
corresponding critical temperature $T_{c}={\hbar}/k_{B}\beta_{c}$ (note that
$T_{0}<T_{c}<T_{1}$), producing thus a first order transition.
\section{Finite Temperature Instantons in Field Theory}
   Studying first order phase transitions in the early universe, Linde [5]
suggested an extension of the previously discussed second order transition
scenario (published by Affleck shortly before) to the field theory case.
Due to spherical symmetry in the spatial coordinates, the equation of motion
for the field is now
\begin{equation}
\frac{{\partial}^{2}{\varphi}}{{\partial}{\tau^{2}}} +
\frac{{\partial}^{2}{\varphi}}{{\partial}r^{2}} +
\frac{2}{r}\frac{{\partial}{\varphi}}{{\partial}r} =
\frac{{\partial}V}{{\partial}{\varphi}}\;\;\;,
\end{equation}
where r is the spherical radial distance. We will choose the potential to be
of the form:
\begin{equation}
V({\varphi})=\frac{\lambda}{2}({\varphi}^{2}-{\mu}^{2})^{2} - F{\varphi}\;\;\;.
\end{equation}
As for the boundary conditions, in the $r$ direction the requirement of finite
action leads to $\varphi{\rightarrow}{\varphi}_{+}$   as  $r{\rightarrow}
{\infty}$, while in the $\tau$ direction periodicity alone no longer suffices,
and we have to impose that the configurations at ${\tau}={\pm}\beta/2,\;0$
have zero derivative along the ${\tau}$ direction to ensure that they are
indeed turning points. We thus have
\begin{equation}
\varphi{\rightarrow}{\varphi}_{+}\mbox{ as }r{\rightarrow}{\infty}\;\mbox{,}\;
\;\;{\partial}{\varphi}/{\partial}{\tau} =0\mbox{ at }{\tau}={\pm}{\beta}/2
\;\mbox{, }0\;\;\;,
\end{equation}
where $\beta$ is the period of the solution. The idea is then that the
relevant solutions should follow the same lines as above: at low temperatures
the solution would be a periodic array of widely separated O(4) spherical
bounces (with radius $R$). As we go into temperatures
$T{\geq}(2R)^{-1}$ however we no longer can fit a bounce in
each period and they start merging into one another, forming an
oscillating ``wiggly'' cylinder analogous to the oscillating solutions of the
quantum mechanical case. These wiggles would then subsequently smooth out as we
increase the temperature and eventually disappear at temperatures
$T>>R^{-1}$, providing a smooth transition into the ``cylindrical'' ($\tau$
independent) instanton corresponding to the thermal activation regime. Fig.4
shows this sequence. As we increase the temperature, the transition would
be then a second order one.

   As noted in the introduction however, Garriga [12] has shown that at least
in the thin wall approximation this is not the case. In this approximation, the
parameter $F$ in the potential above (and correspondingly the energy difference
between the two different phases, $\epsilon=2F$) is very small, and as a
consequence the width
of the wall separating the bubble of true vacuum from the false vacuum is very
small compared to its radius $R$ (i.e., one needs a larger bubble to
compensate for the wall energy). In Lorentzian time then, the dynamics of the
bubble can adequately be described by an effective action (in 3 spatial
dimensions)
\begin{equation}
S = -{\sigma}{\int}{d^{3}}{\xi}{\sqrt{\gamma}}+{\epsilon}{\int}dVdt\;\;\;,
\end{equation}
where the first term is the Nambu action proportional to the area of the world
sheet of the wall ($\sigma$ being the wall tension, $\xi^{a}$ a set of
coordinates on the world sheet and $\gamma$ the determinant of the world sheet
metric), while the second is the volume of the bubble times the energy
difference between the two phases integrated over time.
   With a flat background geometry, adopting spherical coordinates and a
spherical ansatz for the world sheet means that we can write $S$ as
\begin{equation}
S = -{\sigma}{\cal S}_{2}\left[{\int}r^{2}(1-\dot{r}^{2})^{1/2}dt -
\frac{\epsilon}{3{\sigma}}{\int}r^{3}dt\right]\;\;\;,
\end{equation}
where $\dot{r}=dr/dt$ and ${\cal S}_{2}$ is the surface of the unit
sphere. Since the Lagrangian does not depend on time, the energy
\begin{equation}
E = p_{r}\dot{r}-L=\sigma{\cal S}_{2}
\left[\frac{r^{2}}{(1-\dot{r}^{2})^{1/2}}-\frac{\epsilon}{3\sigma}r^{3}\right]
\;\;\;,
\end{equation}
must be conserved. With a little algebra this can be written as
\begin{equation}
\dot{r}^{2}+V(r,E)=0\;\;\;,
\end{equation}
where
\begin{equation}
V=\left[\frac{E}{{\sigma}{\cal S}_{2}}r^{-2} +
\frac{\epsilon}{3{\sigma}}\right]^{-2} - 1\;\;\;,
\end{equation}

   The instantons relevant to the problem are then found by performing a Wick
rotation in (17), whence we obtain
\begin{equation}
\dot{r}^{2}-V(r,E)=0\;\;\;,
\end{equation}
where now $\dot{r}=dr/d\tau$. There are two types of solutions to (19)
depending on the value of $E$: for $E=0$ and $\beta{\rightarrow}\infty$, the
solution is just Coleman's bounce with radius $R=3\sigma/\epsilon$, and action
\begin{equation}
S_{T}=\sigma{\cal S}_{3}R^{3}/4=\sigma{\pi}^{2}R^{3}/2\;\;\;.
\end{equation}
Keeping $E=0$ but introducing a finite
value for $\beta$ produces then a periodic array of bounces in the $\tau$
direction. These solutions exist for as long as $T{\leq}T_{\ast}=(2R)^{-1}$
and will continue to have the same action $S_{T}$, since outside $R$ the field
decays exponentially.
   For $E>0$ though, the inverted potential -V takes the form of a well and the
motion of $r$ in it acquires two turning points, $r_{min}$ and $r_{max}$. The
solution is then wiggly cylinder shaped. If we keep increasing $E$, $-V$
becomes shallower and shallower, and the two turning points start approaching
each other until
they eventually merge at values $E_{0}$,$r_{0}$. From the condition $V(r_{0},
E_{0})=V'(r_{0},E_{0})=0$ ($V'=dV/dr$), we get values for the radius of the
cylinder
\begin{equation}
r_{0}=\frac{2}{3}R\;\;\;,
\end{equation}
and its corresponding energy
\begin{equation}
E_{0}=\frac{4\pi{r_{0}}^{2}}{3}\sigma=\frac{16\pi}{27}{\sigma}R^{2}\;\;\;.
\end{equation}
Since it is static, the solution $r=r_{0}$ will exist at all temperatures, and
its Euclidean action
\begin{equation}
S_{0}=E_{0}\beta\;\;\;
\end{equation}
will be proportional to the period ($=\mbox{temperature}^{-1}$). This will
therefore be the dominant solution at high temperatures.

   It suffices now to compare $S_{T}$ and $S_{0}$ to see that there will be a
temperature for which $S_{T}=S_{0}$, namely
\begin{equation}
T_{c}=\frac{32}{27\pi}R^{-1}\;\;\;.
\end{equation}
At this temperature then, the curves for $S_{0}$ and $S_{T}$ will cross each
other, and consequently the slope of the curve that follows the minimum action
$S_{min}$ will change discontinously. The transition between the two regimes is
thus first order. It can be shown (see Ref.12) that the wiggly cylinder
solutions
 that exist for $0<E<E_{0}$, do, as a matter of fact, always have
greater actions than either the periodic array of bounces or the static
cylinder, and therefore never dominate $S_{min}$.

   The question that we aim to answer is then the following: Is this situation
general for bubble nucleation, or will we  find a second order transition as
suggested by Linde when we go into the thick wall case?.
\section{Algorithm}
   We start with the Euclidean action
\begin{equation}
S_{E}=4\pi{\int}d\tau{\int}dr r^{2}\left [\frac{1}{2}(\dot{\varphi}^{2} +
{\varphi}'^{2}) + \frac{\lambda}{2}({\varphi}^{2}-{\mu}^{2})^{2} -
F\varphi\right]\;\;\;,
\end{equation}
where $\dot{\varphi}=\partial\varphi/\partial\tau$, $\varphi'=\partial\varphi/
{\partial}r$.
   We then rescale the Lagrangian
\begin{equation}
\varphi{\rightarrow}\varphi/{\mu}\;\;\;,\;\;\;\tau{\rightarrow}
{\tau}\sqrt{\lambda{\mu}^{2}}\;\;,\;\;\;r{\rightarrow}r\sqrt{\lambda{\mu}^{2}}
\;\;\;,\;\;\;f=F/(\lambda{\mu}^{3})\;\;\;,
\end{equation}
to get
\begin{equation}
S_{E}=\frac{4\pi}{\lambda}{\int}d\tau{\int}dr r^{2}\left[\frac{1}{2}
(\dot{\varphi}^{2} +
{\varphi}'^{2}) + \frac{1}{2}({\varphi}^{2}-1)^{2} - f{\varphi}\right]\;\;\;.
\end{equation}
The only adjustable parameter in the Lagrangian is now $f$, so by covering its
whole range we should be covering all relevant cases. The equations of motion
are now
\begin{equation}
\frac{{\partial}^{2}{\varphi}}{{\partial}{\tau^{2}}} +
\frac{{\partial}^{2}{\varphi}}{{\partial}r^{2}} +
\frac{2}{r}\frac{{\partial}{\varphi}}{{\partial}r} =
\frac{{\partial}V}{{\partial}{\varphi}}=2({\varphi}^{3}-{\varphi})-f\;\;\;,
\end{equation}
and the boundary conditions are the usual ones
\begin{equation}
\varphi{\rightarrow}{\varphi}_{+}\;\mbox{as}\;r{\rightarrow}{\infty}\;\;,\;\;
\;\;
{\partial}\varphi/{\partial}\tau =0\;\;\mbox{ at }\;\;\tau={\pm}\beta/2
\mbox{,  }0\;\;\;.
\end{equation}
   To solve this equation we have used a multigrid algorithm. Multigrid
methods were first introduced in 1970s by Brandt [13] (see also [14] for a
brief
introduction and further bibliography on the subject), and the basic underlying
idea is the following: let's suppose that we have to solve a linear equation of
the form
\begin{equation}
L_{h}(q_{h})=0\;\;\;,
\end{equation}
where $L$ is some discrete form of the linear differential operator, $q$ is the
discrete solution, and the subscript $h$ denotes a discretization of the
independent variables (i.e., $\tau$ and $r$ in our case) in a mesh with
meshsize $h$. In order to do this, iterative methods usually start with an
approximate solution $\tilde{q}_{h}$ such that
\begin{equation}
L_{h}(\tilde{q}_{h})=d_{h}\;\;\;,
\end{equation}
where $d_{h}$ is called the $defect$. We then look for approximate solutions,
$\tilde{v}_{h}$, of
\begin{equation}
L_{h}(v_{h})=-d_{h}\;\;\;,
\end{equation}
to form
\begin{equation}
\tilde{q}_{h}^{new}=\tilde{q}_{h}+\tilde{v}_{h}\;\;\;.
\end{equation}
$\tilde{q}_{h}^{new}$ will then be closer to the true discrete solution $q_{h}$
than what $\tilde{q}_{h}$ was. Under some general conditions for the form of
the
matrix of coefficients of the discretized algebraic equations, iteration of
this procedure is a convergent process, converging upon $q_{h}$.
   The heart of the matter is then what approximation to use to solve (32) and
 find $\tilde{v}_{h}$.

   As an example we have relaxation methods, where what one
does is to replace $L_{h}$ in (32) by a $simpler$ operator (such as just its
diagonal part for instance), and then solve the set of equations. The drawback
here however is that, as solving the set of discretized equations is basically
a point by point procedure, convergence for the long wavelenght modes of the
solution will be rather slow usually.

   Multigrid methods operate by {\em coarsifying} rather than {\em simplifying}
the problem. Instead of solving the problem in a mesh of size $h$ with a
simpler operator, we {\em project} the problem into a coarser grid of size $H$,
solve there for $\tilde{v}_{H}$ and then {\em interpolate} back to find the
corrections $\tilde{v}_{h}$ in the finer grid. Appart from having to solve far
less equations than in the finer grid, transfering to a coarser grid has the
added advantage of dramatically improving the convergence rate of the long
wavelenght modes.
   However, if applied as such, this process will not only not converge, but
slightly diverge for short wavelength modes. What one does then is to apply one
or more realaxation sweeps before and after transfering to the coarse grid, to
smooth out the short wavelength componenets of the error that relaxation
handles so well.

   Several developments on this basic scheme are now possible. In first place,
it is obvious that nothing prevents us from adding more grids to the algorythm,
going each time to coarser and coarser meshes. In principle nothing prevents us
in fact from going to grids so coarse that we only have to solve a handful of
equations. Starting from such a grid and computing in it not just the
correction, but the full solution $\tilde{q}_{H}$, and then going up the ladder
and so on, eliminates the need for having an initial guessed solution fed into
the algorythm. This is known as the full multigrid method.

   Another possible expansion leads to an algorythm that can solve non-linear
problems. In this case, if we have to solve
\begin{equation}
L_{h}(q_{h})=f_{h}\;\;\;,
\end{equation}
where $f_{h}$ is some right-hand side term, we need a smooth correction
$v_{h}$ to the current solution $\tilde{q}_{h}$ so that
\begin{equation}
L_{h}(\tilde{q}_{h}+v_{h})=f_{h}\;\;\;.
\end{equation}
To do this note that
\begin{equation}
L_{h}(\tilde{q}_{h}+v_{h})-L_{h}(\tilde{q}_{h})=f_{h}-L_{h}=-d_{h}\;\;\;.
\end{equation}
Transfering this equation to the coarser grid after the pre-relaxation sweeps
leads to
\begin{equation}
L_{H}(q_{H})-L_{H}({\cal R}_{H}^{h}\tilde{q}_{h})=-{\cal R}_{H}^{h}d_{h}\;\;\;,
\end{equation}
where ${\cal R}_{H}^{h}$ is the restriction operator from gridsize $h$ to
gridsize $H$. That is, in the coarse grid we solve for
\begin{equation}
L_{H}(q_{H})=L_{H}({\cal R}_{H}^{h}\tilde{q}_{h})-{\cal R}_{H}^{h}d_{h}\;\;\;.
\end{equation}
If the approximate solution is then $\tilde{q}_{H}$, then the coarse grid
correction is
\begin{equation}
\tilde{v}_{H}=\tilde{q}_{H}-{\cal R}_{H}^{h}\tilde{q}_{h}\;\;\;,
\end{equation}
and the corrected fine grid solution will then be given by
\begin{equation}
\tilde{q}_{h}^{new}=\tilde{q}_{h} + {\cal I}_{h}^{H}\tilde{v}_{H}\;\;\;,
\end{equation}
where ${\cal I}_{h}^{H}$ is now the interpolation operator between gridsize $H$
and gridsize $h$. Note that in general ${\cal I}{\cal R}\neq1$, and so
$\tilde{q}_{h}^{new}\neq{\cal I}_{h}^{H}\tilde{q}_{H}$.\\
\\
   For our problem we used full weighting injection for the projection operator
(see Ref.14) and second order polinomial interpolation (fourth order for the
thin wall case). We were precluded from using a full multigrid approach because
we encountered a bifurcation problem. This problem can appear in cases where
the
non-linear equations have more than one solution compatible with the boundary
conditions, and in practice it means that for certain values of the mesh size
$h$ the coefficients of the discretized equations blow up. One is therefore
confined to using meshsizes smaller than that for which the problem appears,
and does not have acces to a mesh coarse enough to be able to solve the
equations exactly. To solve this problem we limited our algorythm to use only
three meshes. As for the needed initial guess solution, what we did was to
obtain the zero temperature profile of the bounce via a simple shooting
routine (remember that at zero temperature the dominant solution is $O(4)$
symmetryc, and therefore the problem can be reduced to a 1-dimensional
differential equation). We then put it in a two-dimensional form and fed it to
the multigrid algorythm.
After computing the form of the zero temperature solution in this
way we then proceeded to compute finite temperature solutions by continuation
in the temperature (i.e., using the previous solution as the guess for the next
one with slightly higher temperature).

   Since Zebra relaxation turned out to be unstable for our problem, we used
line Newton-Jacobi in the $\tau$ direction for the relaxation smoother
(as increasing the temperature meant in practice reducing the mesh size in that
direction).
Special care had to be taken to handle the negative modes of the instantons.
One
of the general conditions that relaxation methods need to converge is the
absence of eigenvalues with opposite signs in the diagonalization of the
matrix of coefficients. Instantons however do have a negative mode in their
perturbation spectrum, as they describe unstable states. Since at any point
relaxation amounts to solving the linearized equations for the correction
around the present value of the solution, a relaxation procedure is bound not
to converge. To take care of this problem what we did was to fix the value of
the field at the center of the instanton, ${\varphi}_{c}$. (What this means in
practice is that after each relaxation
step we multiply the whole solution by a constant so that the center remains at
the desired value without breaking the continuity of the relaxation process
around it). The problem is then equivalent to that of a membrane with the
center and the boundaries fixed to a certain value and subject to a
potential, which should make it stable unless we have choosen
too high a value at the center. In that case, the whole membrane will want to
jump over the potential barrier (i.e., in the language of Ref. 7 we will be
overshooting). Thus, if the value for $\varphi_{c}$ is too high the solution
will
still diverge, but if it is below critical relaxation should converge.
At the boundary
between these two cases should lie then the true value of the instanton at its
center, $\varphi_{Tc}$, and fixing $\varphi_{c}$ to this value should make
the relaxation of the instanton a convergent process.

   We tried this procedure with the
zero and high temperature solutions, where we could get independent figures for
${\varphi}_{Tc}$, with a much higher precision from the shooting routine (since
in both limits the equation of motion turns into a simpler differential
equation). In practice we found that if the value chosen for ${\varphi}_{c}$
was too high
by an amount say $\Delta$ (so that ${\varphi}_{c}-{\varphi}_{Tc}{\sim}\Delta$)
the solution would start first by converging up to an overall accuracy level
roughly
of the order of $\Delta$, and only {\em after} reaching that accuracy level
would
it start to diverge. Thus it usually didn't take long to produce a value for
${\varphi}_{c}$ within a $0.1\%$ difference from the $\varphi_{Tc}$ value
given by the shooting routine, and a correspondingly low level of error for
the whole solution.

\section{Results}
   Using this algorithm then we aim at finding all the relevant finite
temperature
solutions of the equations of motion
\begin{equation}
\frac{{\partial}^{2}{\varphi}}{{\partial}{\tau^{2}}} +
\frac{{\partial}^{2}{\varphi}}{{\partial}r^{2}} +
\frac{2}{r}\frac{{\partial}{\varphi}}{{\partial}r} =
\frac{{\partial}V}{{\partial}{\varphi}}=2({\varphi}^{3}-{\varphi})-f\;\;\;,
\end{equation}
to elucidate whether the transition between the
thermal hopping and the low temperature regimens is first or second order. We
will do this for three different values of the parameter $f$.
   The first value that we will take for $f$ will be chosen so as to check our
procedure against Garriga's predictions for the thin wall case. Although in
order to do this it would be desirable in principle to pick a value for $f$ as
small as possible so that we get a wall as thin as possible, practical
considerations limit how small a value for $f$ we can choose. Since in the
thin wall limit
the wall radius is proportional to $f^{-1}$, the smaller the value of $f$ the
larger the mesh that we would need to cover the whole bubble. But at the same
time we need to have the mesh size small enough so that it can adequately
represent the structure of the wall in a smooth way. Clearly then, if we do not
want to have so many points in the mesh that the whole algorithm becomes
unmanageable, there is a limit as to how small we can set the value of $f$. It
turns out however that for our pourposes $f=0.25$ is more than adequate.

   For such a value of $f$, we get a wall about half as thick as the interior
of the bubble. Fig.5 represents the shape of the potential for this case, an
Fig.6 shows several finite temperature solutions.
To compute the action of each solution we used a two dimensional version of the
Romberg integration code that can be found in Ref. 14. When comparing the
results obtained in this way with those provided by the shooting routine for
zero temperature and the thermal hopping regimes, we consistently found a
difference of about $0.2\%$ of the total value. We feel that a conservative
stimate for the error in the value of $S_{E}$ at each point should then be no
higher than $0.5\%$. Fig.7 gives us the $S_{E}$ vs. $T$ dependence. From
it we can deduce an approximate value for the temperature $T_{c}$ at which the
transition takes place of $T_{c}\approx0.0476$ in our adimensional units, which
is in excellent agreement with the results from Garriga. Use of equation (24)
above yields $T_{c}\approx0.0474$, where we have
used a value of $R$ measured from the center of the bubble to the middle of the
wall. The difference between the two values is well within our error stimates,
and the transition is clearly seen to be first order.

   The reason behind such a good agreement with the thin wall results despite
the fact that the wall is only half as thick as the bubble lies in Figs.6b and
6c. In them we can see two complete periods of the solution for $T=0.91T_{c}$
and $T=T_{c}$ respectively. We see that even at $T=T_{c}$ the two spheres are
far from touching each other, and due to the strong exponential decay outside
the wall they almost do not interact with each other at all. In these
conditions the ansatz used by Garriga for obtaining Eqn.23 (namely that the
action of this solution is still the action for the zero temperature bounce) is
obviously still valid, thus the agreement. Finally Fig.6d shows the
thermal hopping solution that becomes dominant at temperatures $T>T_{c}$.

   The second value of $f$ that we shall use will lay somewhere in the middle
of the spectrum of possible values, at $f=0.55$. Fig.8 shows the shape of the
potential for this case, clearly away from the region where the thin wall
approximation is valid. Fig.9a shows the zero
temperature form of the solution (obviously already within
the thick wall regime), and Figs.9b and 9c show two complete periods
for $T=0.9T_{c}$ and $T=T_{c}$. Here we can see some differences from the
previous thin wall case: the two spheres  collide and start merging
into the wiggly cylinder structure, as shown more clearly in the aereal view
depicted in Fig.9d, but at the critical temperature
the solution (which in fact looks more like a ``sierra'' than like a cylinder)
is still quite far away from the thermal hopping profile shown in Fig.9e.
Finally in Fig.10 we see that, although the transition is still first order it
is obviously much less strongly so that in the previous case. The same
considerations about errors apply in this case as above.

   The final value that we picked for $f$ is very close to its upper possible
limit, $f=0.75$ (for $f=0.8$ the double well structure of the potential totally
disappears and we no longer have neither a metastable state at ${\varphi}_{-}$
nor, consequently, any tunneling). The shape of the potential is again depicted
in Fig.11, where we see that in this case the field tunnels over a very
shallow barrier. Fig.12a shows the shape of the zero temperature bounce in
this case, where we can see that the thick wall features of the solution have
been accentuated. Fig. 12b shows again two complete periods at
$T=0.6T_{c}$. We see however in this case that the term proportinal to
$\partial\varphi/{\partial}r$ has affected the sructure of the bounces, that no
longer are spherically symmetric in this case, but stretched in the $\tau$
direction.  The peaks then start to interact ``earlier'' than in the previous
cases, and at $T=0.6T_{c}$ they have already stablished the wiggly cylinder
or ``sierra'' pattern that Linde advanced. This pattern continues to develop in
Figs.12c and d, where we can see how the solutions at $T=0.96T_{c}$ and
$T=0.999T_{c}$ (the latter already inside the margin error for $T_{c}$ itself)
clearly
and smoothly interpolate between the periodic array of bounces and the thermal
hopping solution. The corresponding graph for the action is shown in Fig.13.
The figure shows a typically second order transition structure, with the curve
for the action of the quantum regime merging smoothly up to the second
derivative into the thermal activation curve.
\section{Conclusions}
   In this paper we have addressed the issue of whether the transition from the
thermal hopping regime to the quantum regime for bubble nucleation is first or
second order, in the context of field theory $\varphi^{4}$ phase transitions.
Our main conclusion is that, whereas for most of the possible values of the
parameter $f$ that breaks the degeneracy of the two vacua the transition is
first order, second order transitions will still occur when the potential
barrier is sufficiently shallow.

   As mentioned in the introduction, any realistic model of the field phase
transition will have to incorporate a temperature dependence in the potential,
the specifics of which will depend on the particular values of the parameters
chosen for the model.
It seems clear from our results however that any model that does not settle
sufficiently rapidly in the very shallow barrier regime will most likely
present a first order transtion between the thermal and the zero temperature
regime, for only very shallow
 barrier penetration produces the very thick bubbles that seem to be needed in
order to have smooth merging of the instantons and a second order transition.
Indeed, were we to expect the results from the
quantum mechanical case to carry over into the the field theory problem, one
could have made the point that this actually had to be the case: since in the
latter situation the point at which
the field exits the barrier after tunneling is still quite far away from the
stable vacuum and relatively close to the metastable state, the field is not
sensitive to the details of the potential beyond the barrier, and close to the
true vacuum, while tunneling. In this case then one could approximate the
potential by either a $-\varphi^{2}+\varphi^{3}$ or a
$-\varphi^{2}+\varphi^{4}$
expression. However, $-q^{2}+q^{3}$ and $-q^{2}+q^{4}$ potentials are well
known to produce second order transitions in quatum mechanics.

   The main practical difference between the two cases will lay in how abruptly
the quantum tunneling rate will depart from the thermally induced one in the
region close to the transition temperature, first order transitions marking a
much
sharper deviation from the thermal hopping regime than second order ones. In
the context of
condensed matter physics the same instanton techniques that we have applied
here have yielded predictions for decay rates of metastable states via
macroscopic quantum tunneling of the system at low temperatures, and about the
nature of the transition from this to the thermal activation regime, that are
amenable to experimental check (i.e. [15], [16-18] among others, and more
recently [19]). In this context we are likely to observe very thick wall
transitions, since in most of the situations of interest in experiments we
either do not have a double well structure in the potential, or have rather
shallow barriers. Indeed, second order transitions have been observed in
Josephson
junctions, and new experiments are being prepared to study the formation of
bubbles of inversed magnetization in thin films under field inversion (a
process completely analogous to the one that we have studied here only that
in $2+1$ dimensions) to check the predicted second order transition for that
case.

{\bf Acknowledgements}

   A.F. would like to thank J. Garriga and E. Chudnovsky for helpful
discussions and suggestions, P. Shellard and C. Rebbi for insightful
comments on the numerical techniques, and specially A. Vilenkin for his insight
and countless suggestions.
\pagebreak
\section{References}
\frenchspacing
\begin{enumerate}

\item G.Blatter, V.B.Geshkenbein, and V.M.Vinokur, Phys.Rev.Lett. {\bf 66},
3297 (1991); G.Blatter and V.Geshkenbein, Phys.Rev. {\bf 47}, 2725 (1993).
\item B.I.Ivlev, Yu.N.Ovchinnikov, and P.S.Thompson, Phys.Rev. {\bf B44},
7023 (1991).
\item E.M.Chudnovsky, A.Ferrera and A.Vilenkin, Phys.Rev. {\bf B} (in press)
\item E.M.Chudnovsky, A.Ferrera, to be published
\item A.Linde, Nucl. Phys. {\bf B216}, 421 (1983).
\item J.Preskill,A.Vilenkin, Phys.Rev. {\bf D 47}, 2324 (1993).
\item J.S.Langer, Ann. Phys. (N.Y.){\bf 41}, 108 (1967)
\item M.B.Voloshin, I.Yu.Kobzarev, and L.B.Okun, Yad. Fiz. {\bf 20}, 1229
(1974)[Sov. J. Nucl. Phys. {\bf 20}, 664 (1975)].
\item C.Callen and S.Coleman, Phys. Rev. {\bf D 16}, 1762 (1977). For a review
of instanton methods and vacuum decay at zero temperature see, e.g., {\em
Aspects of Symmetry} (Cambridge University Press, Cambridge, England, 1985).
\item I.Affleck, Phys. Rev. Lett. {\bf 46}, 388 (1981)
\item E.M.Chudnovsky, Phys. Rev. {\bf A 46}, 8011 (1992).
\item J.Garriga, Phys. Rev. {\bf D 49}, 5497 (1994).
\item A.Brandt, Mathematics of Computation, {\bf 31}, 333 (1977).
\item W.H.Press,S.A.Teukolsky,W.T.Vetterling and B.P.Flannery, {\em Numerical
Recipes in Fortran} (Cambridge University Press, Cambridge, England, 1992).
\item J.Clark {\em et al.}, Science {\bf 239}, 992 (1988), and references
therein.
\item E.M.Chudnovsky and L.Gunther,Phys.Rev.Lett. {\bf 60}, 661 (1988);Phys.
Rev. {\bf B 37}, 9455 (1988).
\item M.Uehara and B.Barbara, J.Phys.(Paris){\bf 47}, 235 (1986);C.Paulsen $et
al.$,Phys. Lett.{\bf 161}, 319 (1991).
\item J.Tejada $et al.$, Phys. Lett. {\bf A 163}, 130 (1992);J.Phys. {\bf C 4},
L163 (1992).
\item D.Prost et al., Phys.Rev.Lett. {\bf B 47}, 3457 (1993).
\end{enumerate}
\pagebreak
\section{Figure Captions}
Fig.1 Motion of the particle in the inverted well: a tunneling trajectory
between $q_{1}$ and $q_{2}$ under the potential
barrier $V$ corresponds to a periodic motion in the inverted well $-V$ with
turning points $q_{1}$, $q_{2}$. \\
\\
Fig.2 (a)Monotonic dependence of $\beta$ on Euclidean Energy, (b)Euclidean
Action vs. Temperature ($\sim\beta^{-1}$) in this case\\
\\
Fig.3 (a)Nonmonotonic dependence of $\beta$ on Euclidean Energy, (b)Euclidean
Action vs. Temperature ($\sim\beta^{-1}$) in this case\\
\\
Fig.4 Different shapes of the instanton as temperature increases: inside the
shaded regions the field attains its true vacuum value. (a) Zero
Temperature bounce,(b) Periodic array of bounces at $T<2R^{-1}$, (c) Wiggly
cylinder, (d) Thermal hopping solution\\
\\
Fig.5 Shape of the potential for f=0.25\\
\\
Fig.6  Instantons describing the nucleation of the bubble for f=0.25.
(a) $T=0$, (b) $T=0.91\,T_{c}$, (c) $T=T_{c}$, (d) $T{\geq}T_{c}$.\\
\\
Fig.7 Temperature dependence of the Euclidean Action: $S_{min}(T)/S(0)$ versus
$T/T_{c}$ for $f=0.25$.\\
\\
Fig.8 Shape of the potential for $f=0.55$\\
\\
Fig.9  Instantons describing the nucleation of the bubble for f=0.55.
(a) $T=0$, (b) $T=0.9\,T_{c}$, (c) and (d) $T=T_{c}$, (e) $T{\geq}T_{c}$.\\
\\
Fig.10 Temperature dependence of the Euclidean Action: $S_{min}(T)/S(0)$ versus
$T/T_{c}$ for $f=0.55$.\\
\\
Fig.11 Shape of the potential for $f=0.75$\\
\\
Fig.12  Instantons describing the nucleation of the bubble for f=0.75.
(a) $T=0$, (b) $T=0.6\,T_{c}$, (c) $T=0.90\,T_{c}$, (d) $T=0.96\,T_{c}$,
(e) $T=0.999\,T_{c}$\\
Fig.13 Temperature dependence of the Euclidean Action: $S_{min}(T)/S(0)$ versus
$T/T_{c}$ for $f=0.55$.\\
\\
\end{document}